# Sensing Remote Bulk Defects Through Resistance Noise in a Large Area Graphene Field Effect Transistor

Shubhadip Moulick, Rafiqul Alam, Atindra Nath Pal*

*Department of Condensed Matter and Materials Physics, S. N. Bose National Centre for Basic Sciences, Sector III, Block JD, Salt Lake, Kolkata 700106, India*

*Email. atin@bose.res.in

**Abstract**

Substrate plays a crucial role in determining transport and low frequency noise behavior of graphene field effect devices. Typically, heavily dope $Si/SiO_2$ substrate is used to fabricate these devices for efficient gating. Trapping-detrapping processes closed to the graphene/substrate interface are the dominant sources of resistance fluctuations in the graphene channel, while Coulomb fluctuations arising due to any remote charge fluctuations inside the bulk of the substrate are effectively screened by the heavily doped substrate. Here, we present electronic transport and low frequency noise characteristics of large area CVD graphene field effect transistor (FET) prepared on a lightly doped $Si/SiO_2$ substrate ($N_A \approx 10^{15} cm^{-3}$). Through a systematic characterization of transport, noise and capacitance at various temperature, we reveal that remote $Si/SiO_2$ interface can affect the charge transport in graphene severely and any charge fluctuations inside bulk of the silicon substrate can be sensed by the graphene channel. The resistance (R) vs. back gate voltage ($V_{bg}$) characteristics of the device shows a hump around the depletion region formed at the $SiO_2/Si$ interface, confirmed by the capacitance (C) – Voltage (V) measurement. Low frequency noise measurement on these fabricated devices shows a peak in the noise amplitude close to the depletion region. This indicates that due to the absence of any charge layer at $Si/SiO_2$ interface, screening ability decreases and as a consequence, any fluctuations in the deep level coulomb impurities inside the silicon substrate can be observed as a noise in resistance in graphene channel via mobility fluctuations. Noise behavior on ionic liquid gated graphene on the same substrate exhibits no such peak in noise and can be explained by the interfacial trapping-detrapping processes closed to the graphene channel. Our study will definitely be useful for integrating graphene with the existing silicon technology, in particular, for high frequency applications.

## Introduction

Graphene, a model two-dimensional Dirac material with a highly tunable electrical transport, has emerged as an alternative component in an electrical circuit[1,2]. Electronic properties of graphene are studied mostly in the field effect geometry, where graphene acts as the transport channel and heavily doped silicon substrate (doping $> 10^{20}/cm^3$) is used as a back gate to tune the charge carriers in graphene. As 2D systems are by definition entirely surface-based and thus highly susceptible to external sources of disorder and in particular, graphene/substrate interface can have a significant impact on the device properties and performance[3–5]. It was observed that the interface traps[6], coulomb impurities[7], grain boundaries[5] or interfacial phonons[3] can limit performance of a graphene device. By removing the substrate completely or putting it on a trap free substrate like hexagonal Boron Nitride (hBN), the mobility of graphene may improve by orders of magnitude[8,9]. One can reach the ultra-clean limit to observe strong interaction driven phenomena like Fractional quantum Hall effect[10], Hoftstadder butterfly[11] etc. by encapsulating graphene with hBN and using graphite as a global gate electrode[12]. However, these are still limited to the micron scale devices, mostly by using scotch tape-based method and dry transfer technique[8]. Achieving these in a large area device is still under research. Chemical vapor deposited large area graphene appeared as an alternative, although, the presence of intrinsic defects, grain boundaries and chemical residues during the transfer process limit the device performance significantly[13,14].

Low frequency noise or flicker noise is an important tool to characterize the device performance and understand the effect of time varying disorder on the channel conductance in a field effect device. Resistance fluctuations in graphene is dominated mostly by the interface states, spatially closed to the graphene channel[15–18]. Both number and mobility fluctuations-based models are proposed to explain the noise data in graphene devices. It was shown that fluctuations in the contacts can also play a major role in the noise performance in graphene devices[19]. However, previous studies on noise in graphene mostly focused on the devices made on degenerately doped Si/SiO$_2$ substrate[15,18,20–23], while to incorporate graphene in present day silicon industry, it is important to fabricate graphene devices on a lightly or moderately doped substrate, particularly for high frequency application. Although, there are few transport studies on graphene field effect devices fabricated on a lightly doped substrate[24–27], there is no investigation of low frequency noise in these devices till date. Apart from the technological relevance, it is important to understand the operating region of the device and investigate the role of bulk defects responsible for resistance fluctuations as observed in silicon-based FETs[28–30].

In this article, we report the electrical transport and low frequency noise measurement on a large area graphene field effect transistor fabricated on a lightly doped (p+ doping ~ $10^{15}$ /cm$^2$) Si/SiO$_2$ substrate. By varying the gate voltage, the substrate undergoes accumulation, depletion and inversion region, confirmed by the Capacitance (C) – Voltage (V) spectroscopy. We observe a significant impact of the space charge region formed at the Si/SiO$_2$ interface on the transport and noise behavior in graphene channel. Resistance (R) vs. back gate voltage ($V_{bg}$) characteristics at room temperature shows a hump near the depletion region, followed by an enhanced resistance noise. A detailed temperature dependence measurement reveals that bulk defects in the substrate, mostly in the depletion regions, are indeed responsible for the observed behavior through mobility fluctuations. For further confirmation, we performed transport and noise measurement in a top gated (ionic liquid) graphene device fabricated on the same lightly doped substrate, where typical noise features were observed which can be explained by the carrier number fluctuations between graphene and the interfacial traps.

**Methods**

CVD graphene field effect devices were prepared from the commercially available CVD graphene (Graphenea, USA) on a boron doped (p+) Si/SiO$_2$ substrate (resistivity ~ 10 ohm-cm). The doping concentration is ~ $5.58 \times 10^{15}$/cm$^3$ confirmed from the CV measurement of the substrate in MOS configuration[31] (**SI Figure S3.1**) To fabricate the device, the CVD graphene was first dry-etched using low power oxygen plasma by keeping the flow rate of oxygen at 20 SCCM with an RF power of 8 W for 15 seconds in an ICPRIE system (SENTECH Si500). Electrical contacts were defined using a laser writer (Microtech, model LW405) using a photoresist (AZ1512- HS), followed by deposition of Ti/Au (5 nm/45 nm) in an e-beam evaporation system and lift-off process. The separation between the voltage leads was 70 µm, having a width of 50 µm as shown in **Figure S1a**. Micro Raman spectra show the characteristics G and 2D peak (**Figure S1c**), confirming the monolayer character of the CVD graphene[32].

Four terminal resistance of the device was measured using the standard ac technique with a lock-in amplifier (MFLI, Zurich instrument) by applying a constant current of ~1 µA with carrier frequency of ~ 1.66 KHz. Keithley 2450 source meter was used to apply the back-gate voltage. The noise measurement was performed using the dual channel lock-in amplifier having a built-in data accusation system following the procedure mentioned in Ref [33]. For C-V measurement, we have used an impedance analyzer (MFIA, Zurich instruments) and a low noise floating voltage source (GS200-YOKOGAWA) in two probe configurations by applying an ac bias of 100 mV. Whole measurement was performed in a custom designed cryogenic insert in a high vacuum environment ($10^{-5}$ mbar). The dipstick was cooled down to 77 K by inserting it in a liquid-nitrogen dewar and a Pt-100 thermometer was mounted alongside the device to measure the temperature of the sample. A Lakeshore-340 temperature controller was used to control and measure the temperature with a temperature stability of ± 5 mK during the measurement.

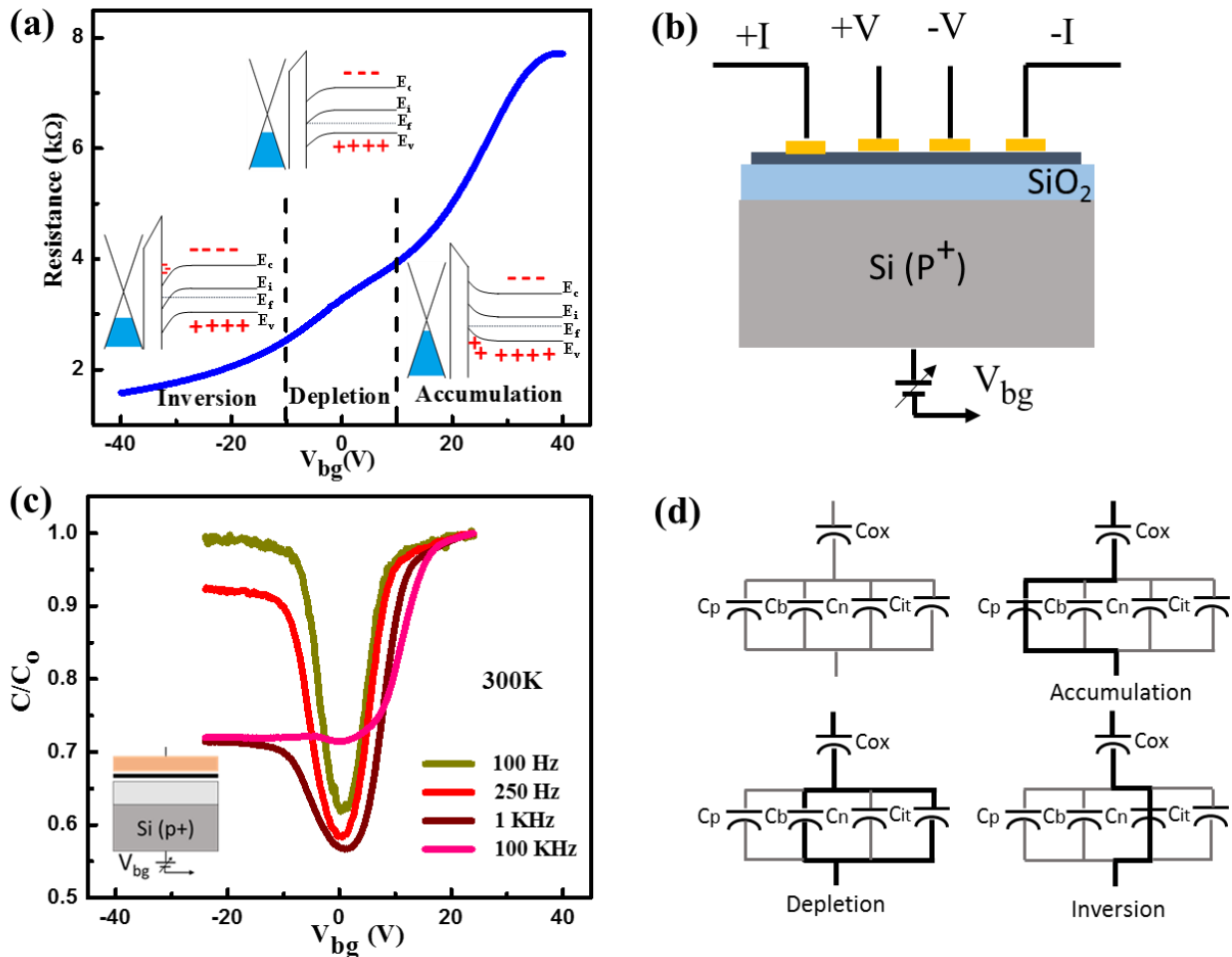

**Figure 1**. **(a)** R-$V_{bg}$ curve of the graphene field effect transistor (solid blue line), showing the charge neutrality point at ~ 38V. Schematic band diagrams of the three characteristics regions: accumulation, depletion and inversion, shown in the respective inset separated by the black dashed lines. **(b)** Schematic of the CVD graphene FET and its measurement scheme. **(c)** C-V characteristics MOS configuration at 300K, for various frequencies. Schematic MOS device structure is shown in the inset. (d) Equivalent capacitance circuit for the MOS device for different substrate condition as a function of gate bias as explained in the text.

**Results and Discussion**

**Figure 1a** (solid blue line) shows the typical variation of four terminal resistance (R) with the global back gate voltage ($V_{bg}$) for the CVD graphene devices grown on lightly doped Si/SiO$_2$ substrate at room temperature (see **Figure 1b** for the schematic of the device). The charge neutrality point (CNP) is ~ 38V as seen from the R-$V_{bg}$ characteristics, indicating that the device was inadvertently p-doped, possibly due to the hole doping caused by PMMA based wet transfer[34–36]. I-V characteristics at different gate voltages were linear (**SI Figure S2.1a**), specifying a good Ohmic contact. Additionally, we observe a hump in the R-$V_{bg}$ characteristics, near $V_{bg}$ ~ 0V. Similar hump was often noticed in low mobility devices either due to contact induced doping or due to the spatial inhomogeneity in the devices[19]. To understand the origin, we checked the R-$V_{bg}$ characteristics in both two and four terminal configurations for different regions of the device (**SI Figure S2.1b**). We have also observed that the hump is present always near zero gate voltage in other fabricated devices and different regions of a single device (**SI Section 7**), pointing out to its different origin.

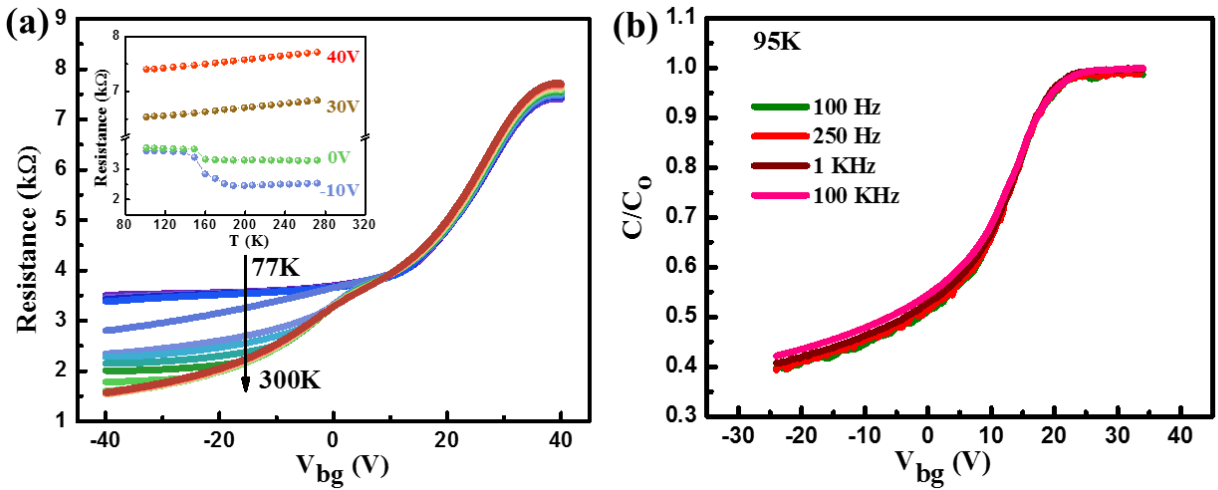

**Figure 2**. **(a)** Temperature dependent R-$V_{bg}$ curve of the graphene-FET device. Inset shows the variation Resistance with temperature for different gate voltages. **(b)** Variation of the normalized capacitance at 95K as a function of gate voltage for different frequency.

To understand the effect of the substrate and various interfaces, we performed C-V measurement by fabricating a MOS capacitor (Metal/Graphene/SiO$_2$/Si/Metal) as shown schematically in the inset of **Figure 1c**. Notably, frequency dependent C-V curves exhibit important characteristics.

Low frequency C-V curve (100 Hz, in **Figure 1c**) clearly depicts the three situations: Accumulation, Depletion and Inversion while varying the gate bias[37]. The region between ~ ± 10V indicates the depletion region, where the capacitance decreases by almost factor of two at $V_{bg}$ ~ 0V. The broad depletion region is possibly due to additional interfacial capacitances that originate due to the trap charges or defects present at the interfaces[38–40]. As expected, no such formation of depletion region was observed in heavily doped substrate (Boron doping, $N_A$~ $1\times10^{20}$/cm$^3$) (**SI Figure S3.3**). At high frequencies (> 100 kHz), the minority electrons cannot follow the change in gate voltage and inversion layer does not form. As the depletion region forms, the effective thickness of the gate dielectric increases resulting in a decrease in gate capacitance, which consequently decreases the ability to add carriers to the graphene channel. This is reflected as a flattening of the R-$V_{bg}$ curve in that region. Similar behavior was observed by Pierre et.al[24] and explained using the drift-diffusion model[41], where a thinner (~ 20 nm) SiO$_2$ dielectric layer was used. Both quantum capacitance of graphene and depletion capacitance play a crucial role in the transport behavior. As the SiO$_2$ layer is thick (~ 300 nm), the role of quantum capacitance[42] is negligible compared to the oxide and depletion capacitances in our case The spread in the C-V characteristics with frequency at 300 K (**Figure 1c**) also indicates the significant effect of the interfacial traps[37,39,43]. This can be understood more clearly from the equivalent circuit of the MOS capacitor shown in **Figure 1d.** The total capacitance can be modelled as the combination of oxide capacitance $C_{ox}$, in series with four parallel capacitances, $C_p$, $C_b$, $C_n$ and $C_{it}$, which appear due to hole charge density, electron charge density, bulk space charge density, and interfacial trap charge density, respectively[40]. As in the accumulation (inversion) region, $Q_p$ ($Q_n$) is the dominant charge carriers $C_p$ ($C_n$) becomes significantly large approaching short-circuit and hence, the equivalent capacitance effectively equals to the oxide capacitance, $C_{ox}$. In the depletion region, however, the bulk and interface capacitance dominate as both $Q_n$ and $Q_p$ becomes significantly small, giving rise to the overall capacitance as $C_{ox}$ in series with the parallel combination of $C_b$ and $C_{it}$ (**Figure 1d**).

Upon decreasing temperature, the hump in the R-$V_{bg}$ characteristics remains in the same position. However, the effect of the gate weakens in the negative side and becomes almost non-responsive below T ~ 150 K (**Figure 2a**). This is more evident from the temperature dependent resistance data at different gate voltages, shown in the inset of **Figure 2a**. In the positive side (accumulation region), resistance decreases with temperature at all gate voltages, indicating metallic behavior. The C-V characteristics at 95 K for different frequencies (**Figure 2b**) show that the device does not go to the inversion region in the negative gate voltage, rather goes into the deep depletion due to the non-availability of minority carriers as a result of carrier freeze out[25]. Moreover, the spread in the C-V curves is reduced by varying the frequencies, indicating that the reduction in the interfacial trapping-detrapping processes at low temperature[37,40].

To get further insight about the transport behavior of the graphene field effect transistor, we investigated the low frequency resistance fluctuations or noise of the device by varying gate voltage and temperature. Noise in our CVD graphene device was measured using four-probe ac method in a constant current mode. For measuring noise, a carrier frequency of 1.66 kHz was used and the time-dependent output of the lock-in amplifier is digitized, followed by multistage decimation of the signal to remove impacts of higher harmonics of the power line or other undesired frequencies, and finally calculation of the power spectral density (PSD)[33]. To avoid heating and other nonlinearities an excitation of 1 µA was used. The linear dependence of the noise PSD over the square of voltage drop across the sample verified the Ohmic nature of the contacts (**SI Figure S4a**). Thermal noise measurements of standard resistors were used to calibrate the setup before actual measurements.

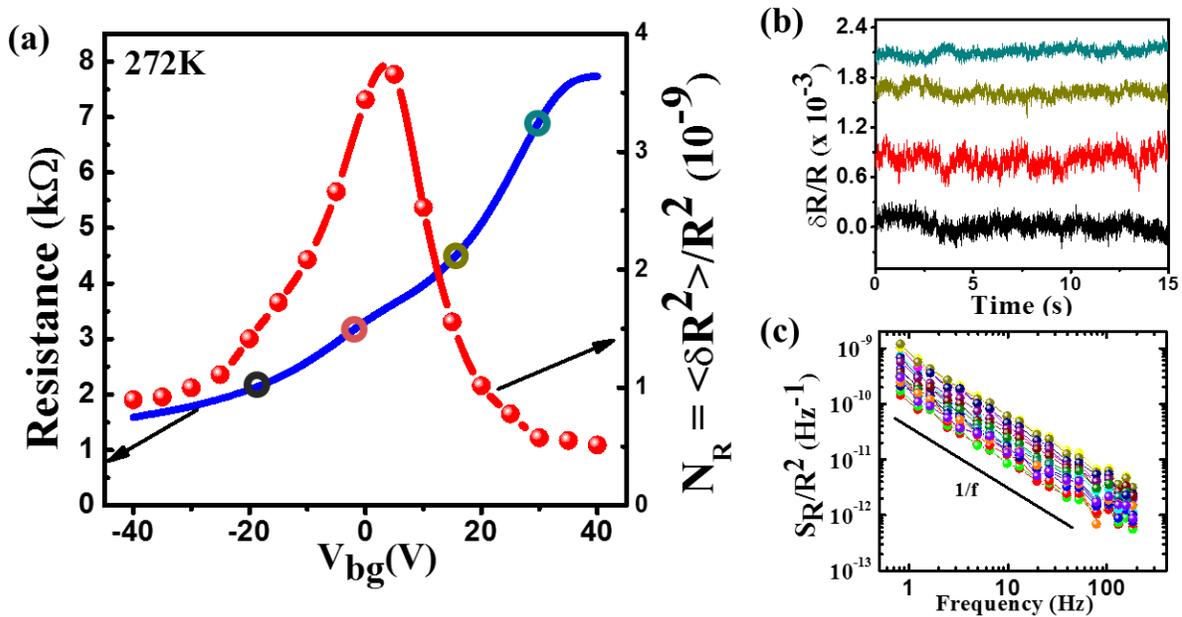

**Figure 3: (a)** Variation of normalized noise PSD, integrated over the measured frequency bandwidth ($N_R$) (red data points) and Resistance (blue solid line) with $V_{bg}$. A clear increase in noise is observed around 0V. **(b)** Time domain resistance fluctuation at different $V_{bg}$, as marked in the R-$V_{bg}$ curve in (a). **(c)** Typical noise power spectra $S_R/R^2$ at various $V_{bg}$, showing 1/f characteristics.

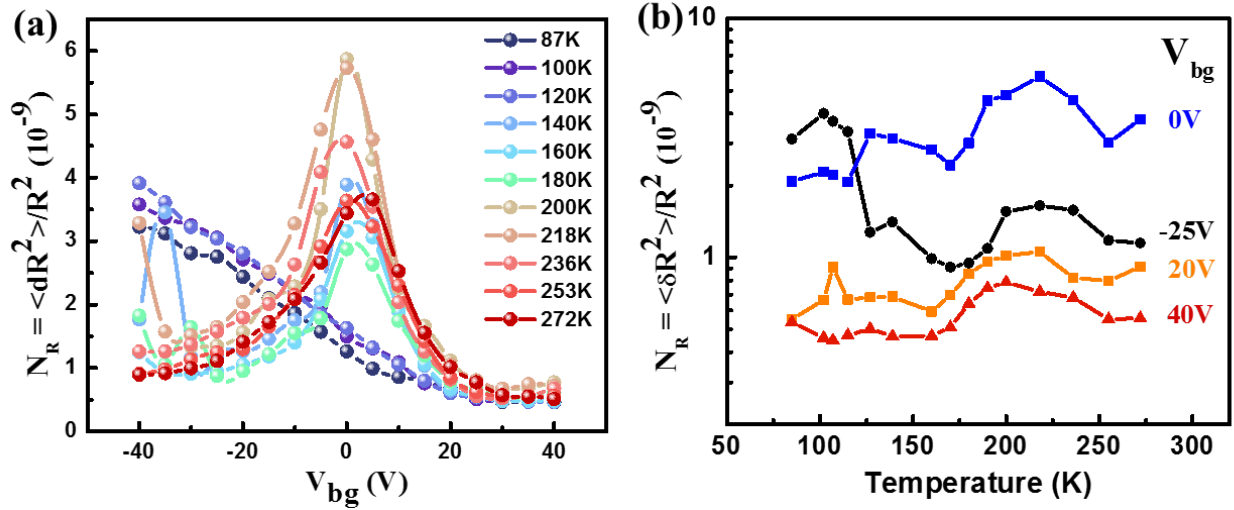

**Figure 4**: **(a)** Variation of normalized noise PSD ($N_R$) with $V_{bg}$ for different temperature **(b)** Temperature dependence of $N_R$ for different fixed $V_{bg}$.

Typically, the normalized power spectral density (PSD) of resistance noise follows the empirical Hooge's relation, $S_R(f) = \frac{\gamma_H R^2}{nA_G f^\alpha}$, where $A_G$ is the area of the sample between the voltage probes, n is the carrier density, $\gamma_H$ is the empirical Hooge's parameter. The noise PSD is proportional to $1/f^\alpha$ with α varies from 0.9-1.1 (**SI Figure S4b**). Here instead of focusing on the amplitude of noise power at a particular frequency or $\gamma_H$, we analyzed the total variance of resistance fluctuation,

$$N_R = <\delta R^2>/R^2 = \frac{1}{R^2}\int S_R(f)df \dots\dots\dots(1)$$

where the numerical integration is carried out over the experimental bandwidth[44]. For noise measurement in our sample, time varying resistance was recorded keeping the gate voltage at a fixed value. The normalized noise PSD ($N_R$) of the device as a function of gate-voltage is shown in **Figure 3a**. The peak in the normalized noise around 0V is confirmed by the time series data of $\delta R/R$ (**Figure 3b**). We have observed that $S_R/R^2 \propto 1/f^\alpha$, with α varying between 0.9 – 1.2 (**SI Figure S4b**), for all applied gate voltages (**Figure 3c**). The peak in the normalized noise PSD corresponds to the gate voltage over which the hump in the R-$V_{bg}$ curve is observed. With varying temperature, $N_R$ vs $V_{bg}$ curve shows the peak around 0 V down to ~ 150 K, and below ~ 150 K the peak vanishes and $N_R$ keeps on increasing with decrease in $V_{bg}$ as shown in **Figure 4a.**

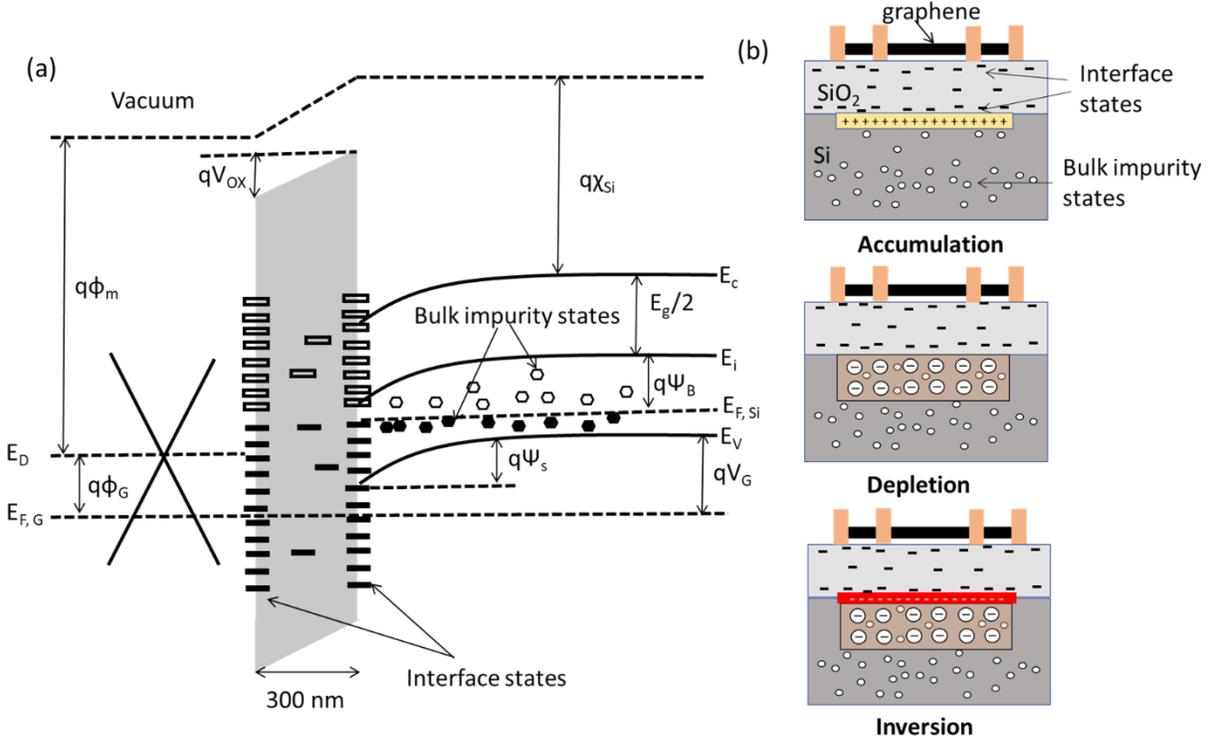

**Figure 5** (a) Schematic energy band diagram of graphene/SiO$_2$/Si (p+) structure with different band parameters, showing the interface and bulk impurity states inside silicon. (b) Schematic diagrams of three different regime obtained by the application of gate voltage. The effect of bulk defects become important only in the depletion region as there no charge layer formed at the SiO$_2$/Si interface to screen any charge fluctuations inside the bulk of the silicon.

Resistance fluctuations in graphene were shown to arise either from the contact or from the trapping-detrapping processes, primarily from the interfacial traps. Two mechanisms were proposed[44]. Firstly, the charge exchange noise ($N_{ch-ex}$), which originates mostly due to the trapping-detrapping processes between graphene and the underlying traps and a quantitative expression was developed based on the correlated number and mobility fluctuations models[15,44–46]. There is another type of noise where no charge exchange will occur, however, fluctuations in the occupancy of the traps may alter the scattering cross-section with time giving rise to fluctuations in resistance based on the local interference model developed by Pelz and Clarke[47], termed as configuration noise ($N_{config}$). Since these two contributions occur independently, total normalized noise magnitude can be written as[44],

$$N_R = N_{ch-ex} + N_{config} = A(n_{it}, T)\left(\frac{d\sigma}{dn}\right)^2 + B(n_{ir}, T)N_c(n) \ldots\ldots\ldots..(2)$$

where $\sigma$ is the conductivity, $n$ is the carrier density of graphene, $n_{it}$ is the density of interfacial traps closed to graphene participating in the charge exchange process, $n_{ir}$ is the density of traps responsible for mobility fluctuations without any charge exchange processes. It is evident that $N_{ch-ex}$ is proportional to the square of the differential mobility ($\sim (d\sigma/dn)^2$) and the density-independent proportionality constant, $A$, will be proportional $n_{it}$ and follows the temperature ($T$) dependence according to the McWhorter's model[48]. The density dependent part of the second term, $N_c(n) = (|n|/n_\Delta)^\gamma)$ and 1 for $|n| > |n_\Delta|$ and $|n| < |n_\Delta|$, respectively, where the parameter $n_\Delta$ denotes the characteristic density below which the charge distribution in graphene is inhomogeneous due to the presence of charge puddles. The parameter $\gamma$ varies between $-1$ and $-2$ for single layer graphene and signifies the enhanced screening ability with increasing density.

The dip in noise at the CNP along with the weak increase with $n$ are typical characteristic of low mobility graphene devices where the charge exchange noise originating due to the interfacial traps dominates the second term in our CVD graphene devices[14,44]. The noise peak near zero gate voltage, however, cannot be described by the models solely based on the interfacial traps. Our results indicate that it might be related to the formation of the depletion region, hence, the bulk defects could be responsible for the resistance fluctuations. We depict the picture schematically in **Figure 5** through the energy band diagram of the graphene/SiO$_2$/Si structure under finite gate bias. Under negative gate bias with respect to graphene, the bands bend upward and there is a formation of depletion region (**SI section 6**). Interfacial traps can exist at both graphene/SiO$_2$ and SiO$_2$/Si interfaces and inside the oxide dielectrics[37]. There also exists SRH type defects[49], responsible for generation – recombination (GR) noise inside the bulk of the silicon[29,50–52]. When the substrate is in the accumulation or inversion region, the effect of the bulk defects will be negligible compared to the interfacial traps, as the accumulation or inversion layer will screen any type of such charge fluctuations. In the depletion region, there will be three effects. Firstly, the effective carrier density of graphene changes almost by factor of 2 when the depletion region is maximum, which itself can increase the configuration noise term in **Equation 2** due to reduced screening. Secondly, the graphene channel will feel significant coulomb fluctuations due to GR noise inside the bulk of the substrate, leading to the increase of configuration noise via mobility fluctuations in graphene[53,54]. Finally, the electric field due to the depletion region may also alter the potential of the interfacial traps and enhance the noise[38,55,56]. At this point it is difficult to decouple the contributions from different mechanisms and get a quantitative estimate of the remote defects through the analytical model.

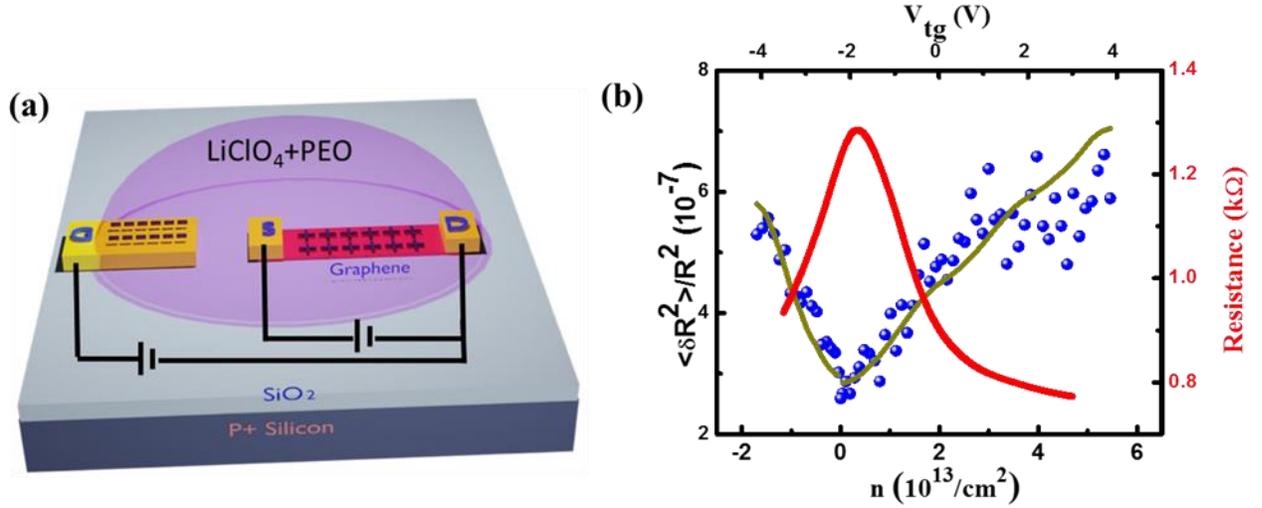

**Figure 6** (a) Schematic of the ionic liquid gated CVD device. (b) Variation of Resistance (red line) and noise (blue dots) as a function of carrier density of liquid ion gated graphene field effect transistor. The green solid line indicates the fitting with the model proposed

Temperature dependent noise data exhibits two important features. Firstly, the noise peak near zero gate voltage vanishes below T ~ 150 K (**Figure 4a**) and secondly, noise at different gate voltages show non-monotonic behavior with temperature[57]. The vanishing of the noise peak at low temperature can be correlated with the carrier freeze out of the silicon substrate. Due to the non-availability of thermally generated minority carriers of the bulk silicon at low temperatures, the inversion layer doesn't form and the $SiO_2$/Si interface goes into deep depletion region. In the freeze out region, the GR noise also decreases as there are not enough carriers available for the trapping-detrapping process in the bulk and hence, the peak disappears. The increase in the noise amplitude with gate voltage in the freeze out region, where the gating ability vanishes, can be explained from the following argument. In the deep depletion region, as there is no formation of inversion layer, the gate electric field cannot be screened effectively, resulting into a lowering of the trap potential near the graphene/$SiO_2$ interface. As a result, noise monotonically increases with gate electric field due to the enhanced noise from the interfacial traps. Temperature dependent normalized noise PSD for different gate voltages are plotted in **Figure 4b**, which clearly shows that noise increases non-monotonically with temperature and becomes maximum in the temperature region 200 K – 230 K. In the inversion region ($V_{bg}$ = -25V, **Figure 4b**), a sudden increase in noise with lowering temperature below T = 150 K is observed apart from the maximum at ~ 220 K. Previous temperature dependence of the noise studies on graphene prepared on a heavily doped substrate exhibit activated behavior with temperature[44], indicating that trapping detrapping processes at the interface plays a significant role following Dutta & Horn model[58]. The non-monotonic behavior of the noise in our case provides another indication that the effect of bulk substrate is more relevant compared to the interfacial traps. As the charge fluctuations inside the substrate is connected to the mobility of the charge carriers, which behaves nonmonotonically with temperature, can be the possible origin of observed noise behavior in our graphene device.

To establish our claim further, we performed noise measurement on a top-gated graphene device, fabricated on the same wafer where an ionic liquid, $LiClO_4$+ PEO matrix, was used as a top gate dielectric, as schematically shown in **Figure 6a**. In an ionic liquid, whenever a potential is applied there is a formation of Debye layer of thickness ~ 1-2 nm at the graphene/liquid interface that helps to induce large number of charge carriers by applying a small gate voltage[59] (see also **SI section 5** for details). **Figure 6b** shows the Resistance (R) vs. top gate voltage ($V_{tg}$) characteristics of the top gated device, exhibiting CNP at ~ -1.9V (**Figure S5a**). We performed the noise measurement on the device and the result is shown in **Figure 6b** (blue circles). There is no peak observed near zero gate voltage as observed in the back gated device. The normalized noise ($N_R$) as a function of gate voltage exhibits a minimum at the CNP point and increases with gate voltage on both the side, while almost saturates at high density in the positive side. As reported earlier[44,60], similar dip in the noise amplitude was observed both in the top gated and the back gated graphene field effect devices. The density dependent noise data can be fitted with the equation, $N_R = A(T)(\partial\sigma/\partial n)^2 + C(T)$, where the first term dominates and depicts the contribution from the charge exchange noise due to the interfacial traps. The density independent second term may arise due to the contact noise or configuration noise as explained in previous reports[60,61].

**Conclusion**

In conclusion, we provide new experimental study on electronic transport and low frequency noise characteristics of a large area CVD graphene field effect transistor fabricated on a lightly doped silicon substrate (with a doping concentration ~ $10^{15}$/cm$^3$). The resistance vs. gate voltage characteristics shows a hump near zero gate voltage at room temperature. By performing C-V measurement in a MOS configuration, we confirm that the hump in resistance appears as a result of the reduced gate capacitance due to the formation of depletion region at the $SiO_2$/Si interface. This has more dramatic effect in the low frequency noise behavior near $V_{bg}$ = 0V, where a large noise peak is observed. By performing temperature dependent noise, along with capacitance and resistance measurement down to 77K, we reveal that this noise is directly related to the formation of the depletion region. It is proposed that the both bulk charge fluctuations in the depletion region, as well as modulation of the energy level of the interface traps may give rise to mobility fluctuations in graphene channel. We also provide additional measurement on a top gated device showing usual behavior which can be explained by charge exchange processes between graphene and the interfacial traps closed to the surface. Through this study we clearly demonstrate that graphene field effect transistor is susceptible to the remote bulk charge fluctuations in a lightly doped substrate, which is not only important for incorporating graphene with the existing silicon technology, but provides an important fundamental understanding on the impact of the remote interfaces on the electron transport in graphene.

**Acknowledgement**

S.M. acknowledges the CSIR Fellowship program, Ministry of Science & Technology, Government of India, for providing research fellowship (09/575(0118)/2017-EMR-I). R.A. acknowledges the INSPIRE Fellowship program (award number IF170926), DST, Government of India, for providing research fellowship. The authors acknowledge the characterization facilities from the TRC project, the clean room fabrication facilities of SNBNCBS.

**Supporting information contains**:

Device characterization, Additional electrical transport data, C-V characteristics, Additional noise data, Liquid ion gated device characterization, Energy band diagram of the MOS structure.

(61) Tersoff, J. Low-Frequency Noise in Nanoscale Ballistic Transistors. *Nano Lett.* **2007**, *7* (1), 194–198. https://doi.org/10.1021/nl062141q.
16

# Supporting Information

**Sensing Remote Bulk Defects Through Resistance Noise in a Large Area Graphene Field Effect Transistor**

Shubhadip Moulick, Rafiqul Alam, Atindra Nath Pal*

*Department of Condensed Matter and Materials Physics, S. N. Bose National Centre for Basic Sciences, Sector III, Block JD, Salt Lake, Kolkata 700106, India*

*\*Email. [atin@bose.res.in](mailto:atin@bose.res.in)*

**Contents**

**Supplementary Note 1:** Device characterization.

**Supplementary Note 2:** Additional transport data.

**Supplementary Note 3:** Additional C-V data.

**Supplementary Note 4:** Additional noise data.

**Supplementary Note 5:** Characterization of the liquid ion gated graphene device.

**Supplementary Note 6:** Energy band diagram of the MOS device.

**Supplementary Note 7:** Additional measurements to confirm the origin of hump.

## 1. Device characterization

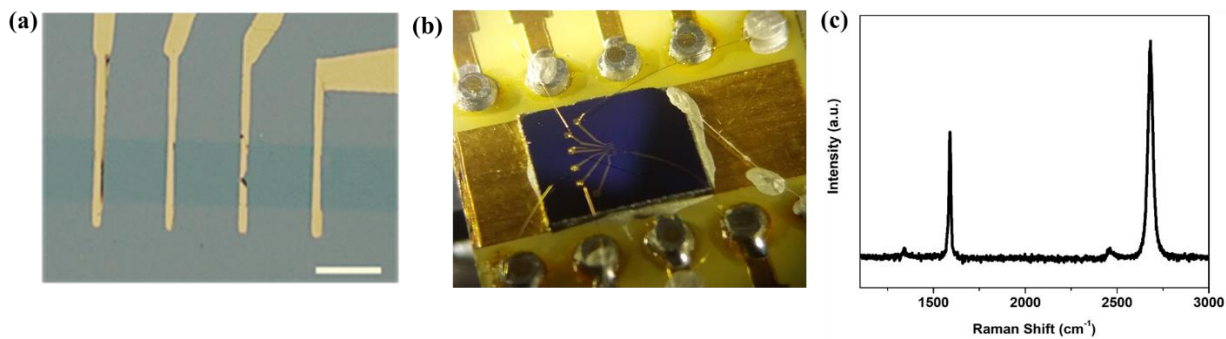

**Figure S1** (a) Optical micrograph of the etched graphene device. The scale bar is 50μm. (b) Optical image of the bonded device. (c) Raman spectra of the CVD graphene exhibiting single Lorentzian peak around 2680 cm$^{-1}$, which confirms the single layer characteristics.

## 2. Additional transport data

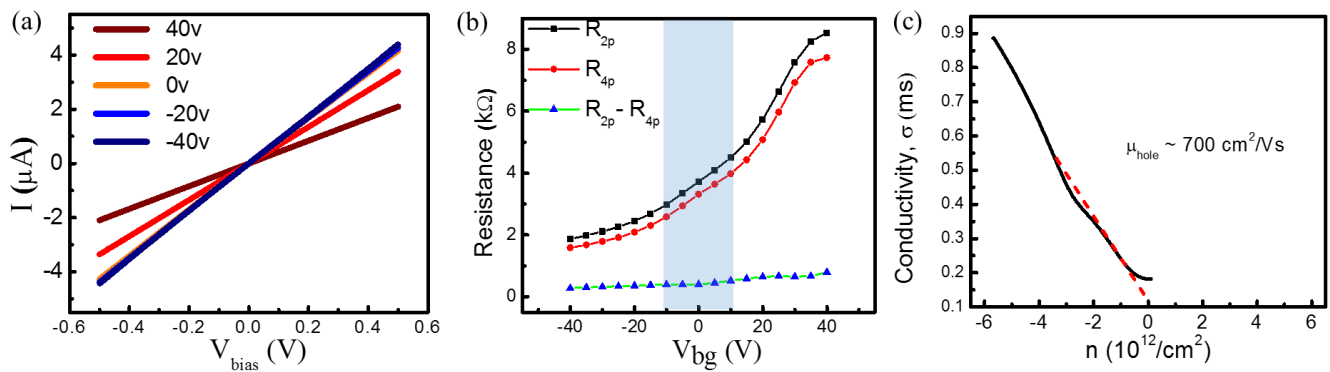

**Figure S2.1** (a) I-V characteristics of the device at different gate voltages show a linear behavior confirming a good Ohmic contact. (b) The variation of resistance in two-probe (black) and four-probe (red) and their difference (blue) with gate voltage. (c) Conductivity (σ) vs. carrier density (n) at T = 272K.

### 2.1. Estimation of contact resistance

To get the contact resistance, we have measured the resistance in both two-probe and four-probe configurations, as shown in **Figure S2b.** The hump near in the resistance $V_{bg} = 0V$ is present in both the configurations, which signifies that the hump originates from the graphene channel. The variation of the contact resistance with the gate voltage is shown in **Figure S2b**, by subtracting four probe resistance ($R_{4p}$) from two probe resistance ($R_{2p}$)[1].

## 2.2. Calculation of mobility

The mobility of a graphene device can be calculated from Drude model using the relation,

$$\mu = \frac{\sigma}{ne}$$

Where, σ is the conductivity, e is the electronic charge, and n is the charge carrier density.

In our device, it is difficult to estimate the exact carrier concentration using the parallel plate capacitor model due to the formation of depletion region, where the capacitance is bias dependent. Hence, for simplicity we calculate the hole mobility from the σ vs n curve only in the accumulation region. In the accumulation region, the capacitance is approximately equal to the oxide capacitance ($1.5 \times 10^{-8}$ F/cm$^2$). By linear fitting the small region near the Dirac point (**Figure S2c**) the hole mobility was found to be~700 cm$^2$/Vs. Due to the possibility of gate leakage at higher gate voltages, we limited our measurement only till ± 40V.

## 3. C-V characteristics

### 3.1. Estimation of doping concentration of the silicon substrate from C-V measurement

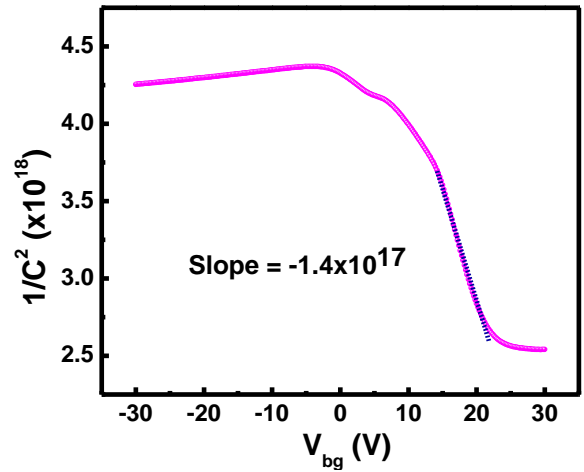

**Figure S3.1**. $1/C^2$ vs $V_{bg}$ curve for the MOS capacitor at 300K by applying an ac bias of 100 mV of frequency of 100 kHz.

The used silicon substrate has a resistivity of 1-10 Ω-cm, which corresponds to a doping concentration of $1 \times 10^{15}$ cm$^{-3}$ according to the specifications (https://www.graphenea.com/collections/buy-graphene-films/products/monolayer-graphene-on-sio2-si-4-wafer). C-V characteristics at high frequency are often used estimate the doping concentration as following [2]. The relation to calculate doping concentration is given by,

$$N_{sub} = \frac{2}{q\varepsilon_s A^2 \left(\frac{\Delta 1/C^2}{\Delta V_{bg}}\right)}$$

Where, q is the electronic charge, $\varepsilon_s$ is the permittivity of the substrate, A is the area of the gate. As shown in the main text (Figure 1c, inset), for C-V measurement, we have fabricated a CMOS structure on the lightly doped substrate with an effective area of the gate equal to 0.1225 cm². From the slope of the $1/C^2$ vs $V_{bg}$ graph (shown in Figure S3.1.) the doping concentration can be estimated as ~ $5.58 \times 10^{15}$ cm$^{-3}$.

### 3.2. Temperature dependent C-V characteristics

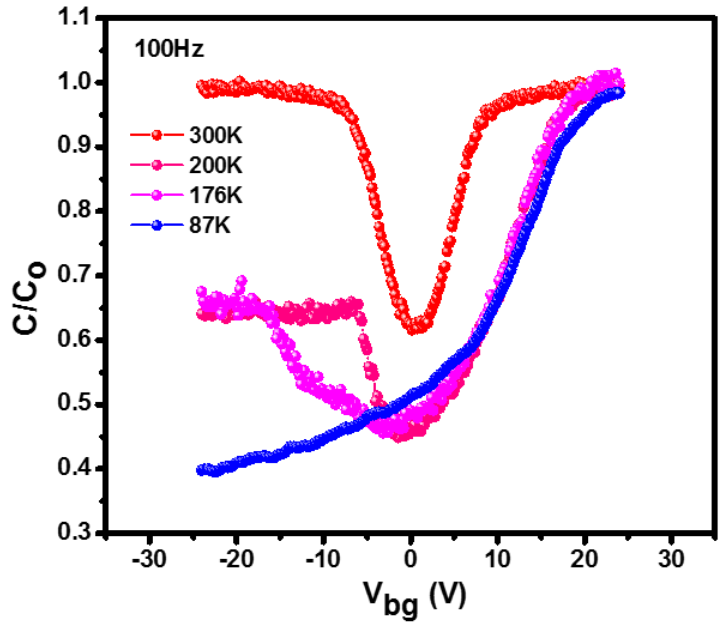

**Figure S3.2** Temperature dependent C-V characteristics of Si/SiO$_2$/Graphene/Au MOS capacitor at 100Hz.

The variation of capacitance with gate voltage at 100Hz for different temperature is shown in **Figure S3.2**. The red curve corresponds to the C-V curve at 300K, where three characteristics regions are clearly visible. The capacitance in the accumulation and in the inversion regions are equal to the oxide capacitance. By lowering the temperature, we observe that the inversion capacitance cannot reach the oxide capacitance because of the non-availability of sufficient thermally generated minority carriers. When the temperature is below the freeze out temperature (~ 150K), the depletion region keeps on increasing and goes into the deep-depletion, where the gate loses its capability of inducing charge carriers in graphene.

### 3.3 C-V characteristics of highly doped substrate

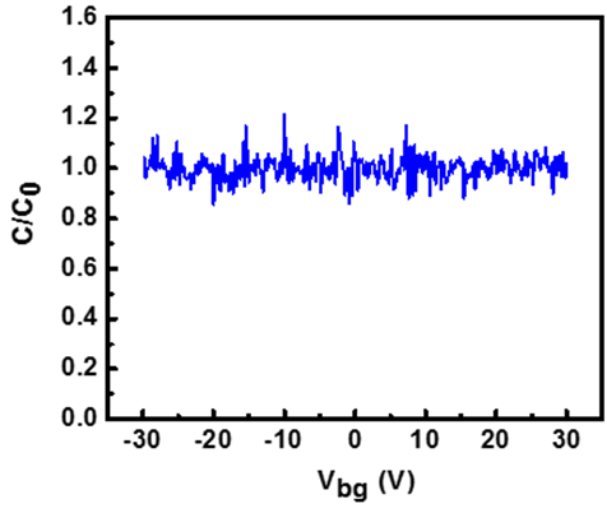

**Figure S3.3** variation of capacitance with gate voltage for a highly doped silicon measured in MOS geometry.

Figure S3.3 shows the variation of capacitance with gate voltage for a heavily doped substrate ($N_A \sim 10^{20}$ cm$^{-3}$) in MOS capacitor geometry, as schematically shown in the inset of **Figure 1c** in the main manuscript. As depletion region is negligible, the capacitance remains almost constant throughout the gate voltage range.

### 4. Additional noise data

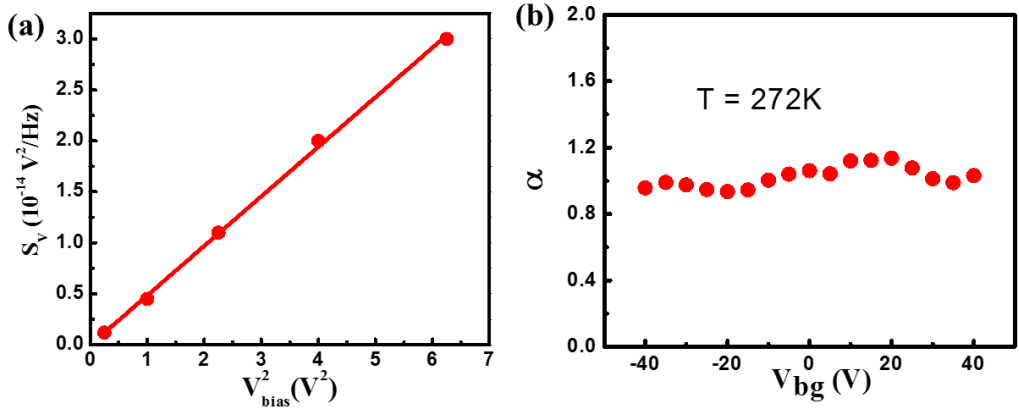

**Figure S4**. (a) Variation of noise power spectral density, $S_v$ with $V_{bias}^2$, exhibiting linear behavior expected for Ohmic contact. (b) Variation of α with $V_{bg}$, calculated from the linear fit of the 1/f noise data.

### 5. Liquid ion gated device characterization

To isolate the effect of the depletion layer at the Si/SiO$_2$ interface, we have measured transport and noise in a top gated (ionic liquid) device prepared on the same substrate. CVD graphene device was prepared using optical lithographic technique. After etching a desired area of graphene metal electrodes (Ti/Au 5/60 nm) were deposited using e-beam evaporation. Separate large area electrode was made alongside the graphene channel to provide bias to the liquid ion gate. Top gating was achieved by using solid polymer electrolyte consisting of LiClO$_4$ and polyethylene oxide (PEO) in the ratio 0.12:1. The polymer acts as a dielectric material. When a top gate voltage, $V_{TG}$, is applied, the ions move toward the sample and form the Debye layer[3]. For a polymer electrolyte the thickness of the Debye layer is about 1-5 nm which acts as a parallel plate capacitor. Considering the Debye layer thickness of 2nm for our case of lithium percolate + PEO top gate and the dielectric constant of PEO as 5, we get the gate capacitance, $C_{TG}$ = 2.2x 10$^{-6}$ Fcm$^{-2}$, which is much higher than $C_{BG}$. For this reason we can induce huge number of charge carriers by applying a small gate bias. In general, the doping concentration at each gate voltage is given by,

$$V_{TG} = \frac{\hbar V_F \sqrt{\pi n}}{e} + \frac{ne}{C_{TG}}$$

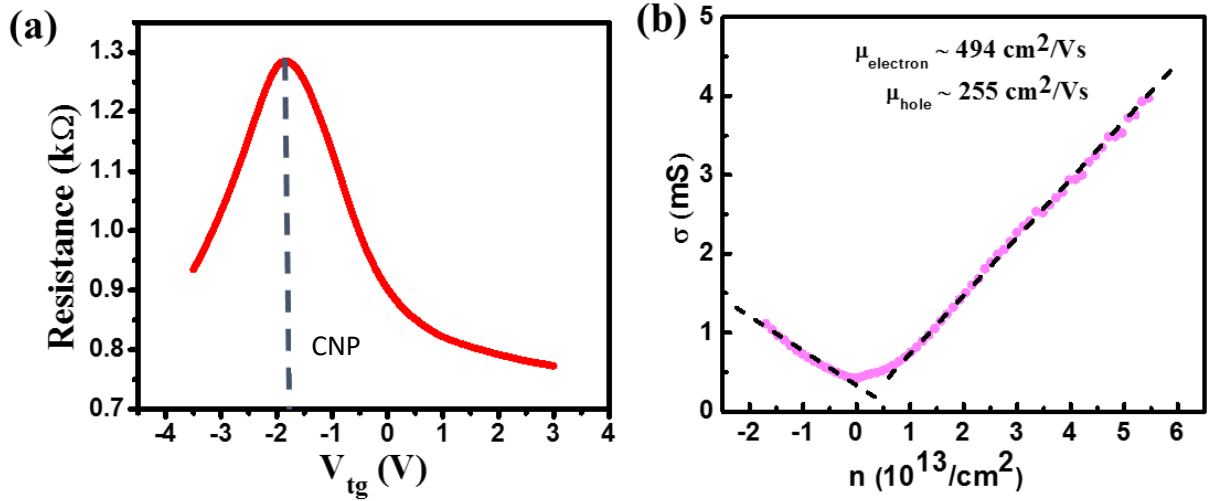

**Figure S5**. (a)The resistance (R) vs top gate voltage (Vtg) are plotted, exhibiting CNP at -1.9 V. (b) conductivity (σ) vs carrier density (n), are plotted for the top gated graphene device (pink dots) The electron and hole mobilities were obtained by the linear fitting of the σ vs. n curve.

Using the $C_{TG}$ = = 2.2 x 10$^{-6}$ F cm$^{-2}$ and $V_F$ = 1.1x10$^6$ m s$^{-1}$, we get

$$V_{TG}(volts) = 1.16 * 10^{-7}\sqrt{n} + 0.723 * 10^{13} n$$

Where *n* is in units of cm$^{-2}$.

The R-$V_{TG}$ measurement were carried out in vacuum in two probe configurations. DC voltage is applied to the top gate using KITHLEY 2450 source meter. A waiting time of 5 to 10 seconds was used before taking any data to stabilize the movement of the ions inside the top gate

## 6. Energy band diagram of the MOS device

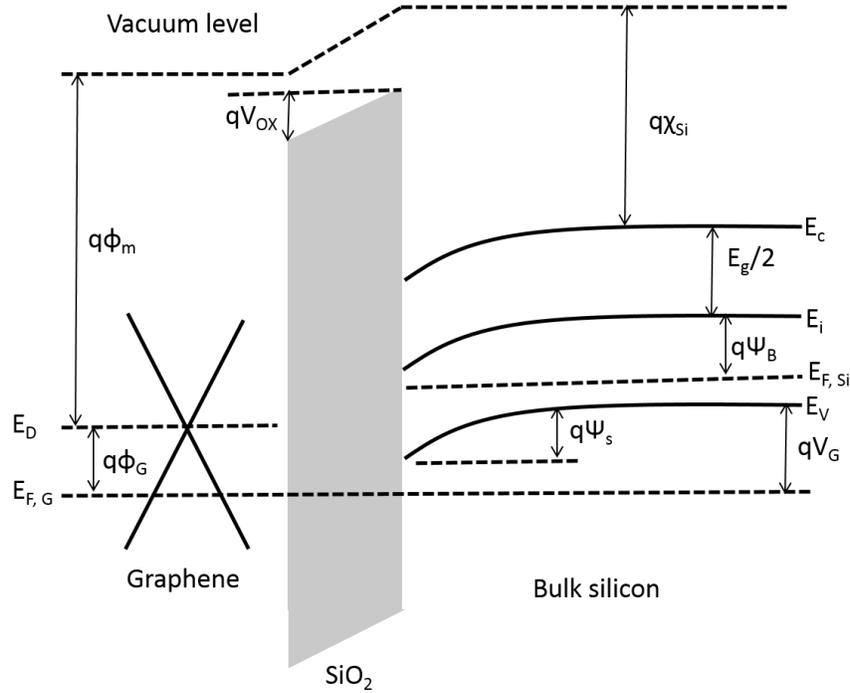

**Figure S6**. Illustration of band bending of the graphene/SiO$_2$/Si interface.

The energy band diagram of the graphene/SiO$_2$/Si/Metal structure is depicted in the **Figure S6**. Whenever a gate voltage (V$_G$) is applied, the energy bands are either pushed upwards or downwards depending on the sign of the gate voltage. As shown in the figure, when a negative gate voltage is applied to the bulk silicon the energy bands will be pushed upward which results into downward bending of the energy bands (q$\psi_s$) at the silicon-silicon dioxide interface. Through the capacitive coupling the Fermi level in the graphene channel shifts (q$\phi_G$) due to the application of gate voltage. This results in to a potential difference across the SiO$_2$ which results into upward tilting (q$V_{ox}$) of the energy bands toward the gate[4,5] From the energy band diagram one can write down the energy conservation equation as

$$-q(V_G - V_{Ch}) + q\psi_B - q\psi_s + \frac{E_g}{2} + q\chi_{Si} - qV_{OX} - q\phi_m + q\phi_G = 0$$

Where, $\chi_{Si}$ is the silicon affinity, $E_g$ is the band gap of silicon, $\phi_m$ work function of graphene, $V_{OX}$ is the voltage drop across SiO$_2$, $\phi_G$ is the graphene channel potential, $\psi_S$ is the surface potential, and $V_G$ is the applied gate voltage. By solving the equation numerically and using the drift diffusion model, it is possible to obtain the transport characteristics in three different regions[4].

## 7. Additional measurements to confirm the origin of hump

### 7.1 R-V$_{bg}$ characteristics of different devices

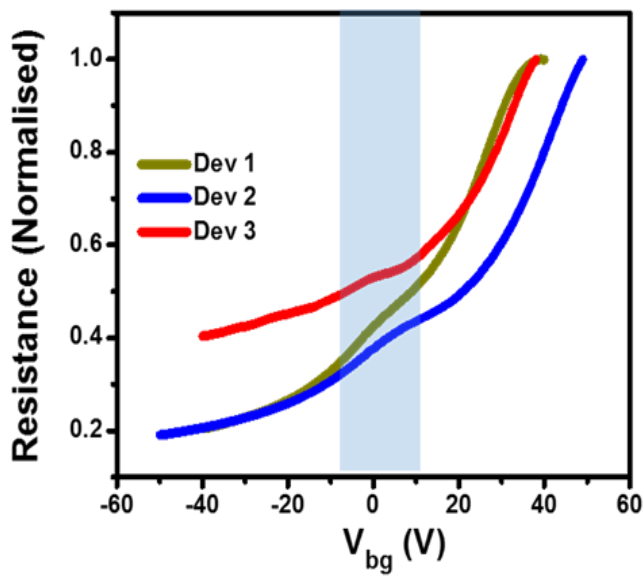

**Figure S7.1**. R-V$_{bg}$ characteristics from three different devices at room temperature exhibiting the hump near V$_{bg}$ = 0V.

### 7.2 Effect of top gate dielectric on the electrical transport behavior

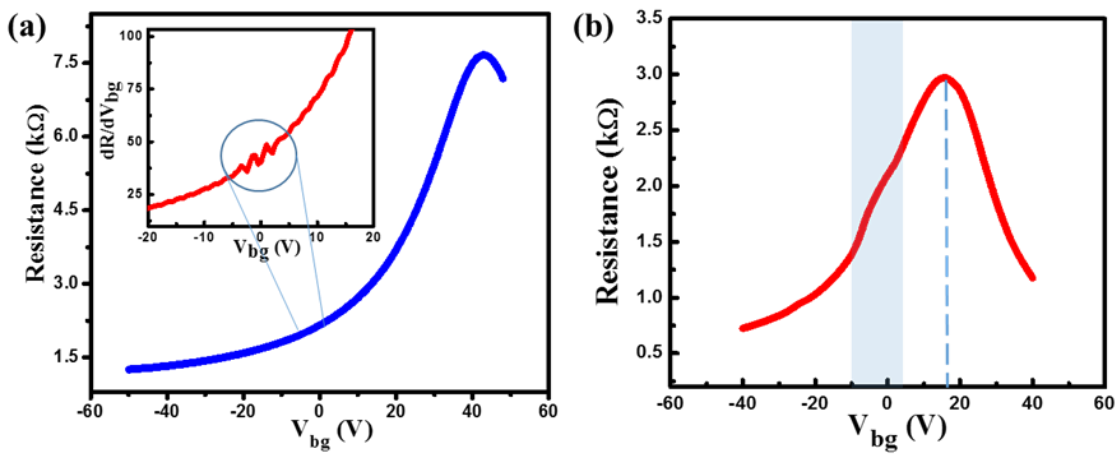

**Figure S7.2.** (a) R-V$_{bg}$ characteristics of a GFET showing charge neutrality point ~ 42V. Inset shows the (dR/dV$_{bg}$) vs V$_{bg}$ curve, which shows small fluctuation around 0V. (b) R-V$_{bg}$ characteristics of the same device after putting the liquid ion gate exhibiting a shift of CNP from ~ 42V to ~ 18V.

In few devices we did not observe the hump near zero gate voltage in the R-$V_{bg}$ characteristics (e.g., **Figure S7.2a**). This is possible when a large number of interfacial traps are present. These interfacial traps provide significant capacitance ($C_{it}$, **Figure 1d,** main manuscript) in the depletion region, because of which we may not be able to observe prominent effect of the depletion capacitance. However, after putting the ionic liquid ($LiCO_4$) in PEO matrix, we observe two effects. Firstly, the CNP shifts to ~ 15V from ~42V and the hump is recovered at zero gate voltage (**Figure S7.2b**). This is due to the neutralization of the interface traps by the ionic impurities and shifts the CNP towards lower gate voltage. Concomitantly, it also reduces the effect interface capacitance which helps to observe the hump more clearly.

**7.3 Transport data on different portion of the device to exclude the inhomogeneity effect**

We show four-terminal R-$V_g$ characteristics from three different regions of a single device (**Figure S7.3**). In all of these configurations, even if the Dirac point is at a different location, the hump always appears around 0V. The fact that the hump appears around 0V in each configurations and in multiple devices, suggests that it is not a result of contact, defects, or inhomogeneity in the graphene sample.

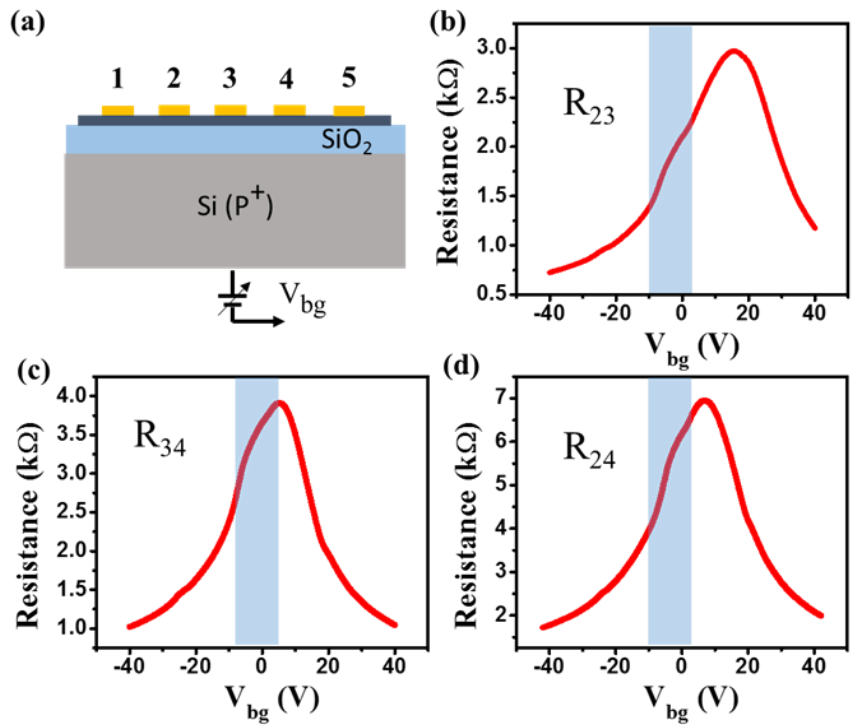

**Figure S7.3**. (a) The schematic of the device, showing the 5 contacts. The current is passed between 1 and 5 and the voltage drops is measured in three different configurations: (b),(c),(d), respectively.

References.

**TOC**

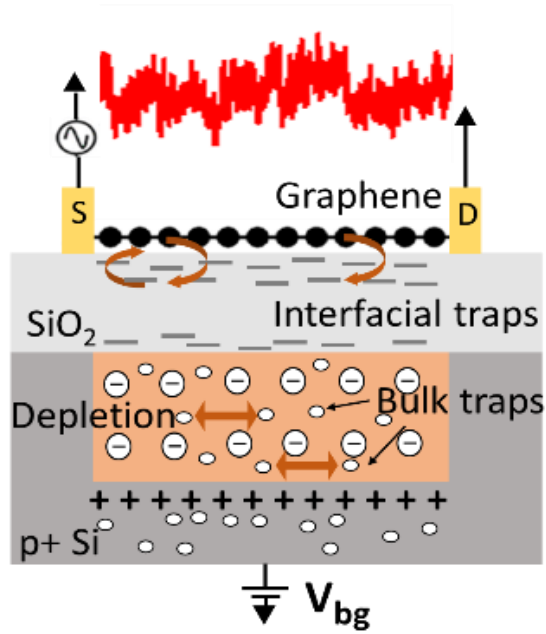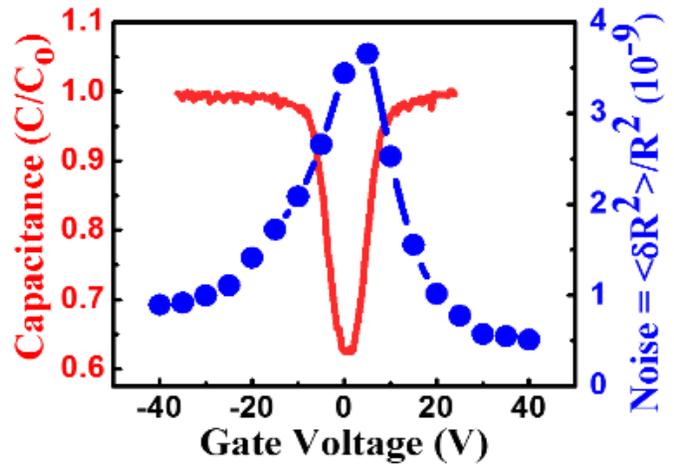